\documentclass[showpacs,prb,superscriptaddress,twocolumn,amsmath,amssymb]{revtex4}

\usepackage{epsfig}
\usepackage{graphicx}
\usepackage{color}

\begin{document}

\title{Entanglement entropy and quantum phase transitions in quantum dots
coupled to Luttinger liquid wires}

\author{Moshe Goldstein}
\affiliation{The Minerva Center, Department of Physics, Bar-Ilan
University, Ramat-Gan 52900, Israel}
\affiliation{Department of Physics, Yale University, 217 Prospect Street,
New Haven, CT 06520, USA}
\author{Yuval Gefen}
\affiliation{Department of Condensed Matter Physics,
The Weizmann Institute of Science, Rehovot 76100, Israel}
\author{Richard Berkovits}
\affiliation{The Minerva Center, Department of Physics, Bar-Ilan
University, Ramat-Gan 52900, Israel}

\begin{abstract}
We study a quantum phase transition which occurs in a system composed
of two impurities (or quantum dots)
each coupled to a different interacting (Luttinger-liquid) lead.
While the impurities are coupled electrostatically,
there is no tunneling between them.
Using a mapping of this system onto a Kondo model, we show
analytically that the system undergoes a Berezinskii-Kosterlitz-Thouless
quantum phase transition
as function of the Luttinger liquid parameter in the leads and the
dot-lead interaction.
The phase with low values of the Luttinger-liquid parameter is characterized
by an abrupt switch of the population between the impurities
as function of a common applied gate voltage.
However, this behavior is hard to verify numerically since one
would have to study extremely long systems. Interestingly though,
at the transition the entanglement entropy drops from a finite value of
$\ln(2)$ to zero. The drop becomes sharp for infinite systems.
One can employ finite size scaling to extrapolate the transition point
and the behavior
in its vicinity from
the behavior of the entanglement entropy in moderate size samples.
We employ the density matrix renormalization group numerical procedure
to calculate the entanglement entropy of systems with lead lengths of up to
480 sites. Using finite size scaling we extract the transition value
and show it to be in good agreement with the analytical prediction.
\end{abstract}


\date{\today}

\maketitle

\section{Introduction}

There has been a recent flurry of activity relating entanglement
entropy (EE) \cite{amico08}
(known also as the von Neumann entropy in quantum physics,
and related to
the Shanon entropy in information theory \cite{aude02,vidal03,calabrese04},
and the Bekenstein-Hawking
entropy in the framework of black holes \cite{srednicki93,fiola94,holzhey94}),
to quantum phase transitions (QPTs)\cite{qpt,vojta06}
in condensed mater \cite{amico08,lehur08}.
The notion of of EE for a many-body system in a pure state arises when one
divides it into two distinct regions: A and B. The
entanglement between the subsystems A and B is measured by the EE $S_{A/B}$
related to $\rho_A$ or $\rho_B$,
the reduced density matrix of regions A or B, respectively.

Specifically, using the Schmidt decomposition, one can express any
many-body pure state of the entire system, $|\Psi\rangle$,
as the sum of two orthonormal
basis sets of regions A ($\{|\phi_{A,i}\rangle\}$) and B
($\{|\phi_{B,j}\rangle\}$), such that
\begin{eqnarray}
|\Psi\rangle=\Sigma_{i}\alpha_{i} |\phi_{A,i}\rangle \otimes |\phi_{B,i}\rangle,
\label{phi}
\end{eqnarray}
with real $\alpha_i \geq 0$ obeying $\Sigma_{i}\alpha_{i}^2=1$.
This basis is closely related to the eigenbasis of the reduced density operators
$\hat \rho_{A/B}={\rm Tr}_{B/A}|\Psi\rangle$, i.e., 
\begin{eqnarray}
\hat \rho_{A/B}=\Sigma_{i}\alpha_{i}^2|\phi_{A/B,i}\rangle \langle \phi_{A/B,i}|.
\label{e3}
\end{eqnarray}
A unique measure of entanglement between the two regions A and B, is the
von Neuman entropy of the reduced density matrix:
\begin{eqnarray}
S_{A/B}=-\Sigma_{i}\alpha_{i}^2\ln(\alpha_i^2), 
\label{e4}
\end{eqnarray}
equivalent to the Shannon entropy of the squared 
Schmidt coefficients $\alpha_i^2$. This measure is also called
the EE. Evidently $S_A=S_B$.

\begin{figure}
{\epsfxsize=3in \epsffile{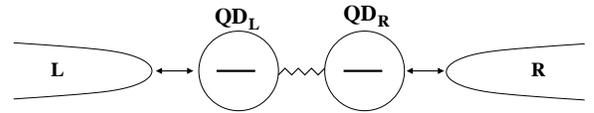}} 
\caption{The system:  two single state quantum dots
(impurities)
each of which is
coupled to a different 1D lead.
Electrons may hop from the left (right) lead to the left (right) dot,
but not between the dots. Thus,
the two dots are coupled only by an electrostatic interaction.
}
\label{system}
\end{figure}

Much effort was 
devoted to establishing the connection between EE and QPTs in
many-particle one-dimensional (1D) systems.
It has been shown that in the critical regime of these models there are
deviations from the
celebrated area law \cite{srednicki93},
which states that the EE $S_A$ should depend only
on the surface area between regions A and B,
hence it is constant (i.e., independent of system size) for 1D systems.
This is indeed the case when the system has a finite correlation
length (when it is gaped \cite{hastings04}). For cases in which
the correlation length is infinite (i.e., in the critical regime)
a logarithmic correction appears in the EE with a universal prefactor.
Using conformal field theory arguments
the EE of an infinite one-dimensional
Luttinger Liquid (LL)\cite{giamarchi03} and various corresponding
spin chains was calculated,
and the universal prefactor was related to the
central charge of the underlying conformal field theory
\cite{holzhey94,vidal03,calabrese04}.
Corrections to the EE due to
finite size \cite{zhou06,laf06}
and the presence of a defect
\cite{zhao06,levine04,peschel05,igloi09,ren09,eisler10}
were also considered.
For example,
a static impurity embedded in a spinless LL, no matter how weak it is,
will result (for repulsive electron-electron interactions
\cite{mattis74,kane92})
in effectively
severing the sample at the impurity location at low energies,
leading to a vanishing ground state conductivity.
It is therefore not unexpected that this is manifested in the behavior
of the EE, which tends to vanish for an infinite LL with a static impurity
\cite{levine04}. Dynamical impurities, e.g.,
an impurity with a resonant state
which may fluctuate between an occupied and a vacant 
configuration, may on the other hand lead to
a different behavior.
Understanding the effect of dynamical impurities on transport
\cite{kane92,furusaki93,auslande01,postama02,polyakov03,komnik03,nazarov03,lerner08,goldstein10a,elste10} 
as well as on thermodynamic \cite{furusaki02,sade05,lehur05,weiss0607,wachter07,weiss08,fiete0810,goldstein09,goldstein10b,goldstein10d,hamamoto10,mora10,goldstein10c}
properties of 1D LLs, has recently garnered much interest.

In the present paper we define three tasks. The first is to analytically
demonstrate that the system depicted in Fig.~\ref{system},
two impurities (quantum dots, QDs), each
coupled to a different external lead, which are intercoupled only
via electrostatic interactions (no inter-impurity tunneling),
undergoes a Berezinskii-Kosterlitz-Thouless
(BKT) QPT \cite{qpt,vojta06} as function
of the LL parameter and the dot-lead interaction.
This transition is similar in nature to  the QPT recently predicted
in the case where a quantum point contact 
is coupled to one of the impurities \cite{goldstein10}. 
The QPT manifests itself
in the behavior of the population of both impurities.
One of the impurities (corresponding to a broad level) swaps its population
with the second (narrow) impurity as a function of a common external
gate voltage applied on the dots.
Similar switching in different contexts have been studied extensively
\cite{hackenbroich97,baltin99a,baltin99b,silvestrov00,berkovits05,sindel05,koenig05,silva02,meden06,golosov0607,karrasch06a,karrasch06b,karrasch07,goldstein07,goldstein08a}. 
By mapping this problem onto that of a multiflavored
Coulomb gas we have recently demonstrated that if the leads are non-interacting
(regular Fermi-Liquids)
the population switching is steep but not abrupt \cite{goldstein10},
in agreement with previous studies
\cite{martinek03,lee07,kashcheyevs07,kashcheyevs09,silvestrov07}.
Once the external leads are LLs, we will show that
this switching becomes abrupt at
a critical value of the electron interactions in the lead (i.e.,
at a critical value of the LL parameter) or between the lead
and the impurity.

The second goal
is to understand the EE behavior in this system, especially its sensitivity
to the QPT. As we shall see, the system we consider here may be mapped
onto an effective Kondo model \cite{hewson}.
When the leads are noninteracting
or weakly interacting (the LL parameter is close to one) the system
corresponds to a anti-ferromagnetic Kondo phase, and therefore
the two disconnected parts of the system are nevertheless entangled,
leading to a finite EE. On the other hand, for stronger interaction
(lower values of the LL parameter) the system is in a
ferromagnetic Kondo phase, in which no entanglement
between the two parts exists (even though they are still correlated),
and therefore the EE is zero. The crossover between
the finite value and zero at the critical interaction
corresponding to the QPT should be sharp for an infinite system.

Finally, as a third goal, we use EE to numerically study the properties of this
QPT. Density matrix renormalization group (DMRG) \cite{white92,dmrg}
is the most convenient numerical
method to calculate the ground state properties
of this interacting system, since EE appears naturally in the procedure.
One might wonder though why not
calculate directly the population of the impurities and see when
population switching becomes discontinuous? This turns out to be extremely
difficult since, in order to distinguish between a real discontinuity and
a sharp transition, one must exceed length scales of order of the
inverse relevant Kondo temperature
\cite{martinek03,lee07,kashcheyevs07,kashcheyevs09,silvestrov07,goldstein10}.
As we shall see, the latter 
is quite small even for noninteracting leads,
and is further suppressed by LL correlations (until it vanishes
at the transition point). One possible solution to this problem
is to use DMRG with soft boundary conditions \cite{bohr06,berkovits11}.
Here we shall show that by combining EE and BKT finite size scaling
it is possible to obtain a reliable estimate of the critical interactions
in the lead for manageable lead sizes.

This paper is organized as follows: In the next section we introduce the
model. Then in Sec. \ref{sec-map} we detail the mapping of
our model onto a Kondo model, and illustrate the appearance of a BKT
QPT. An
analytical calculation of the dependence of the transition point on the
parameters of the system is given. In the following section
(Sec. \ref{sec-num}), we calculate the EE using the DMRG numerical
method, employing three different finite size procedures to determine the
critical value of the interaction in the lead. All the different
procedures give similar estimates of the critical value,
which are in good agreement
with theory in the range of parameters for which the theory is valid.
We conclude with a discussion of the results obtained in
the previous sections (Sec. \ref{sec-dis}).

\section{Model}
\label{sec-mod}

Here we consider a situation in which two single-state QDs 
(impurities) are coupled each to a different 1D wires in the LL regime
(see Fig.~\ref{system}).
The two impurities are coupled only by an electrostatic interaction.
The spin degree of freedom is ignored
(experimentally this corresponds to the presence
of a strong magnetic field or ferromagnetic dots and leads which
polarize the electron spin).
The system is described by the following Hamiltonian:
\begin{eqnarray}
H=H^L_\text{lead}+H^R_\text{lead}+H_\text{impurities},
\label{hamiltonian}
\end{eqnarray}
where the Hamiltonian $H^\ell_\text{lead}$ of lead $\ell$
($\ell=L$, $R$ for left, right, respectively)
is given by
\begin{eqnarray}
H^\ell_\text{lead} =
- t_\ell \sum_{j=1}^{N_\ell-1} c_{\ell,j}^{\dagger} c_{\ell,j+1} +
 \text{H.c.} + \nonumber\\
U_\ell \sum_{j=1}^{N_\ell-1} (n_{\ell,j}-\nu_\ell) (n_{\ell,j+1}-\nu_\ell) ,
\label{lead}
\end{eqnarray}
here $c_{\ell,j}^{\dagger}$ is the creation operator of an electron on the
$j$th site of lead $\ell$ (whose total length is $N_\ell$),
$n_{\ell,j}=c_{\ell,j}^{\dagger} c_{\ell,j}$
is the population of such a site, $t_\ell$ ($U_\ell$)
is the hopping matrix element (interaction)
between the sites in the lead,
and $\nu_\ell$ is the electronic density in the lead.
At half-filling ($\nu_\ell = 1/2$,
which we will always use in the numerical calculations)
the system is in the LL phase
for $-2<U_\ell/t_\ell<2$,\cite{uu}
with the LL interaction parameter $g_\ell$ and the velocity of excitations $v_\ell$
given by
\cite{giamarchi03,g-parameter}:
\begin{eqnarray}
g_\ell &=& \frac{\pi}{2 \cos^{-1}[-U_\ell/(2 t_\ell)]},
\label{gg} \\
\frac{v_\ell}{2 t_\ell} &=&
\frac{\pi}{2} \frac{\sqrt{1-[U_\ell/(2 t_\ell)]^2}}{\cos^{-1}[U_\ell/(2 t_\ell)]}.
\label{vv}
\end{eqnarray}

The impurities Hamiltonian is
\begin{multline}
H_\text{impurities} =
\varepsilon_L n_{L,d} + \varepsilon_R n_{R,d} +
U_{LR} \left( n_{L,d} - \textstyle\frac{1}{2} \right)
\left(n_{R,d} - \textstyle\frac{1}{2} \right)\\
+ U^\prime_{L} \left( n_{L,d} - \textstyle\frac{1}{2} \right) \left( n_{L,1} - \nu_L \right)
+ U^\prime_{R} \left( n_{R,d} - \textstyle\frac{1}{2} \right) \left( n_{R,1} - \nu_R \right)
\\
+ t^\prime_L d_{L}^{\dagger} c_{L,1} + t^\prime_R d_{R}^{\dagger} c_{R,1} + \text{H.c.}
\label{impurity}
\end{multline}
where $d_{L(R)}^{\dagger}$ is the creation operator of an electron on the
left (right) impurity,
and $n_{\ell,d}=d_{\ell}^{\dagger} d_{\ell}$ is the population
of the $\ell$-th dot.
Here $\varepsilon_{L(R)}$ is the left (right) impurity energy
(which may be varied by applying an external gate voltage;
we henceforth assume that the latter has equal effect on both impurities,
$\varepsilon_\ell = \varepsilon^{(0)}_\ell - V_g$), and 
$U_{LR}$ is the Coulomb coupling between the two impurities.
Each impurity is coupled to the corresponding lead by both
a local electrostatic interaction of strength $U^\prime_\ell$,
and a tunneling term parametrized by 
a hopping matrix elements
$t^\prime_\ell$.
The latter gives rise to level broadenings
$\Gamma_\ell = \pi |t^\prime_\ell|^2 \rho_0$, with
$\rho_0$ being the local density of states on the last site
of the lead at the Fermi energy
[which is equal to $1/(\pi t_\ell)$ at half filling].


\section{Mapping onto Kondo}
\label{sec-map}
Our analytic calculations are based on the
Anderson-Yuval mapping to a Coulomb gas \cite{yuval_anderson,schotte71,wiegmann78,cardy81,vladar88,si93,fabrizio95,kamenev_gefen,
goldstein09,goldstein10b,goldstein10,goldstein10d,goldstein_phd}.
In this approach one expands the partition function to all orders
in the dot-lead tunneling matrix elements $t^\prime_L$ and $t^\prime_R$,
and evaluates the resulting correlation functions at $t^\prime_\ell=0$.
Thus, the partition function becomes a sum over all possible
imaginary time histories of tunneling in and out of each level,
and can be cast in the form of a grand-canonical partition function
of classical particles (charges) representing these hopping events.
Because of the dot-lead interaction, each such process involves
Fermi edge singularity physics \cite{noziers69,mahan}, hence is associated
with a Fermi edge singularity exponent
$\kappa^\ell_\text{FES}$,
which is defined through the long-time behavior of the
correlation function
$\langle d_\ell^\dagger(\tau) c_{\ell,1}(\tau)
c^\dagger_{\ell,1}(0) d_\ell (0) \rangle \sim \tau^{-\kappa^\ell_\text{FES}}$
at $t^\prime_\ell = 0$.
In addition, the level width $\Gamma_\ell$ should be replaced
by an effective value $\Gamma^\ell_\text{FES}$,
which includes the prefactors in this correlation function.
The value of $\kappa^\ell_\text{FES}$ plays a crucial role in the following.
It has been studied in our previous works \cite{goldstein09,goldstein10b,goldstein10d};
here we will repeat our main findings.

When $g_\ell=1$ (the lead is noninteracting), we have the
usual resonant level model, for which
$\kappa^\ell_\text{FES} = \left( 1 - \frac{2}{\pi} \delta_\ell \right)^2$ and
$\Gamma^\ell_\text{FES} = \pi \left| t_\ell \right|^2 \rho_0 \cos(\delta_\ell)$,
where $\delta_\ell=\tan^{-1} (\pi \rho_0 U^\prime_\ell/2)$ is the phase
shift experienced by the electrons near the Fermi energy
in lead $\ell$ due to the corresponding dot-lead interaction \cite{noziers69}.
In the general situation ($g_\ell$ not necessarily equal to unity),
standard bosonization treatment yields
$\kappa^\ell_\text{FES}=[1-g_\ell U^\prime_\ell /(\pi v_\ell)]^2/g_\ell$,
$v_\ell$ being the velocity of the bosonic phase excitations in lead $\ell$,
and $\Gamma^\ell_\text{FES} = \pi \left| t^\prime_\ell \right|^2 \rho_0$.
Taking the limit $g_\ell=1$ we see that,
while within a fermionic description $\kappa^\ell_\text{FES}$ is expressed
in terms of the phase shifts $\delta_\ell$, turning to a bosonized
framework these $\delta_\ell$
are replaced
by their leading order dependence on $U^\prime_\ell$.
This is due to the linearization of the spectrum.
Hence, the values of $\kappa^\ell_\text{FES}$ and
$\Gamma^\ell_\text{FES}$
in any particular
model are renormalized by irrelevant operators not appearing in the
Luttinger model.
Boundary conformal field theory arguments show that
$\kappa^\ell_\text{FES}$ is related to finite size corrections to the
spectrum of the lead with different potentials at its ends, 
which may be evaluated by analytical or numerical means.
The results indicate \cite{goldstein09,goldstein10b,goldstein10d} that
in general
$\kappa^\ell_\text{FES} = ( 1 - 2 g_\ell \delta_\text{eff}^\ell / \pi )^2 / g_\ell$ and
$\Gamma^\ell_\text{FES} =
\pi \left| t^\prime_\ell \right|^2 \rho_0 \cos(\delta_{\text{eff}}^\ell)$
for some effective phase shift $\delta_{\text{eff}}^\ell \in [-\pi/2,\pi/2]$
which reduces to the usual phase shift $\delta_\ell$
when the lead is noninteracting.
For our model of the lead, i.e., a tight-binding chain
with nearest-neighbor interactions, one can employ the Bethe
ansatz\cite{g-parameter} to find\cite{goldstein09,goldstein10b,goldstein10d}:
\begin{equation}
\delta_{\text{eff}}^\ell = \tan^{-1} \left[ \frac{U^\prime_\ell}{\sqrt{(2t_\ell)^2-U_\ell^2}} \right].
\label{eqn:alpha_bethe}
\end{equation}
This discussion is summarized in table~\ref{tbl:fes}.

\begin{table}
\caption{\label{tbl:fes}
The Fermi edge singularity exponent and the effective level width.
See the text for further details.
}
\begin{ruledtabular}
\begin{tabular}{cccc}
  & Non-interacting lead & Bosonization & General expression \\ \hline \\[-0.3cm]
$\kappa^\ell_\text{FES} $ & $\displaystyle\left( 1 - \frac{2 \delta_\ell}{\pi} \right)^2$ &
$\displaystyle\frac{1}{g_\ell} \left( 1- \frac{g_\ell U^\prime_{\ell}}{\pi v} \right)^2$ &
$\displaystyle\frac{1}{g_\ell} \left( 1 - \frac{ 2 g_\ell \delta^{\ell}_\text{eff}}{\pi} \right)^2$ \\[0.3cm]
$\Gamma^\ell_\text{FES}$ &
$\Gamma_\ell \cos(\delta_\ell)$ &
$\Gamma_\ell$ &
$\Gamma_\ell \cos(\delta^{\ell}_\text{eff})$
\end{tabular}
\end{ruledtabular}
\end{table}

The Coulomb gas partition function can then be written in a standard form
\cite{cardy81,vladar88,si93}.
The imaginary time history of the system
(which is a circle whose circumference is the inverse temperature $1/T$)
is divided into intervals in which
the system is in one of four possible states of the two dots:
$\alpha=00$, $10$, $01$, and $11$,
corresponding to both dots being empty,
only the left dot being occupied,
only the right dot being occupied, and both dots being populated,
respectively.
The state $\alpha$ has a dimensionless energy $h_\alpha$,
measured in units of $1/\xi$, where $\xi$ is a short-time
(high-frequency) cutoff.
The intervals are separated by hopping events, which are
the classical Coulomb gas particles, as mentioned above.
Their minimal separation is limited to $\xi$,
which is thus of the order of the inverse bandwidth, $\xi \sim 1/t_\ell$.
A transition from configuration $\alpha$ to configuration $\beta$
($\alpha \ne \beta$)
is associated with a fugacity $y_{\alpha \beta} = y_{\beta \alpha}$,
and a two-component vector charge
$\vec{e}_{\alpha \beta} = -\vec{e}_{\beta \alpha}$
(the two components correspond to the left and right lead, respectively),
obeying the triangle rule
$\vec{e}_{\alpha \gamma} + \vec{e}_{\gamma \beta} = \vec{e}_{\alpha \beta}$.
Physically, the components of $\vec{e}_{\alpha \beta}$ represent the
effective change in the charge of each lead
in the corresponding transition.
Values of these parameters for the system discussed
are summarized in Table~\ref{tbl:cg_params}.
The partition function reads: 
\begin{multline}
\label{eqn:cg_cardy1}
 Z = \sum_{N=0}^{\infty} \sum_{\alpha_i}
 y_{\alpha_1 \alpha_2} y_{\alpha_2 \alpha_3}
 \dots y_{\alpha_{N-1} \alpha_N}
 y_{\alpha_N \alpha_1} \times \\
 \negthickspace \negthickspace
 \int_0^{1/T} \negthickspace \frac{\text{d} \tau_{N}}{\xi}
 \int_0^{\tau_N - \xi} \negthickspace \frac{\text{d} \tau_{N-1}}{\xi}
 \dots \negthickspace \negthickspace 
 \int_0^{\tau_3 - \xi} \negthickspace \frac{\text{d} \tau_2}{\xi}
 \int_0^{\tau_2 - \xi} \negthickspace \frac{\text{d} \tau_1}{\xi} 
 \text{e}^{- S ( \{ \tau_i, \alpha_i \})}, 
\end{multline}
where $N+1 \equiv 1$, so that $\tau_{N+1} \equiv \tau_1 + 1/T$.
The classical Coulomb gas action is:
\begin{multline}
\label{eqn:cg_cardy2}
 S( \{ \tau_i, \alpha_i \} ) =
 \sum_{i<j=1}^N \vec{e}_{\alpha_i \alpha_{i+1}} \cdot \vec{e}_{\alpha_j \alpha_{j+1}}
 \ln \left\{ \frac{ \pi \xi T } {\sin [ \pi T (\tau_j-\tau_i) ] } \right\} \\
 + \sum_{i=1}^N h_{\alpha_{i+1}} \frac{\tau_{i+1} - \tau_{i}}{\xi}.
\end{multline}

One can now write down a set of 15 renormalization group (RG) equations
for the Coulomb-gas parameters, which are valid to second order in the fugacities
$y_{\alpha \beta}$ but are otherwise exact
\cite{yuval_anderson,cardy81,vladar88,si93}:
\begin{align}
 \label{eqn:rg1}
 \frac{d y_{\alpha \beta}}{d \ln \xi} = &
 \frac{2-\kappa_{\alpha \beta}}{2}
 y_{\alpha \beta}
 + \sum_{\gamma} y_{\alpha \gamma}y_{\gamma \beta}
    e^{ (h_{\alpha} + h_{\beta})/2 - h_{\gamma} }, \\
 \label{eqn:rg2}
 \frac{d \kappa_{\alpha \beta}}{d \ln \xi} = &
 - \sum_{\gamma} y_{\alpha \gamma}^2 e^{ h_{\alpha} - h_{\gamma} }
    \kappa^{\alpha}_{\beta \gamma}
 - \sum_{\gamma} y_{\beta \gamma}^2 e^{ h_{\beta} - h_{\gamma} }
    \kappa^{\beta}_{\alpha \gamma}, \\
 \label{eqn:rg3}
 \frac{d h_{\alpha}}{d \ln \xi} = & h_{\alpha}
 - \sum_{\gamma} y_{\alpha \gamma}^2 e^{ h_{\alpha} - h_{\gamma} }
 + \frac{1}{4} \sum_{\beta, \gamma} y_{\beta \gamma}^2 e^{ h_{\beta} - h_{\gamma} },
\end{align}
where
$\kappa_{\alpha \beta} \equiv \left| \vec{e}_{\alpha \beta}
\right|^2$ and $\kappa^{\alpha}_{\beta \gamma} \equiv \kappa_{\alpha
\beta} + \kappa_{\alpha \gamma} - \kappa_{\beta \gamma}$. 


\begin{table}
\caption{ \label{tbl:cg_params}
Parameters appearing in the Coulomb gas expansion, 
Eqs.~(\ref{eqn:cg_cardy1})--(\ref{eqn:cg_cardy2}).
The values of the Fermi edge singularity exponents
$\kappa_\text{FES}^\ell$ and the effective level widths
$\Gamma^\ell_\text{FES}$ are summarized in Table~\ref{tbl:fes}.
}
\begin{ruledtabular}
\begin{tabular}{ccc}
 Fugacities & Charges & Energies \\ \hline
 $y_{00,10} = \displaystyle\sqrt{\frac{\Gamma^L_\text{FES} \xi}{\pi}}$ & 
 $\vec{e}_{00,10} = \left(\displaystyle \sqrt{\kappa^L_\text{FES}}, 0 \right)$ &
 $h_{00} = 0$ \\
 $y_{00,01} = \displaystyle\sqrt{\frac{\Gamma^R_\text{FES} \xi}{\pi}}$ & 
 $\vec{e}_{00,01} = \left( 0,\displaystyle \sqrt{\kappa^R_\text{FES}} \right)$ &
 $h_{10} = \varepsilon_L \xi$ \\
 $y_{10,11} = \displaystyle\sqrt{\frac{\Gamma^R_\text{FES} \xi}{\pi}}$ & 
 $\vec{e}_{10,11} = \left( 0, \displaystyle \sqrt{\kappa^R_\text{FES}} \right)$ &
 $h_{01} = \varepsilon_R \xi$ \\
 $y_{01,11} = \displaystyle\sqrt{\frac{\Gamma^L_\text{FES} \xi}{\pi}}$ & 
 $\vec{e}_{01,11} = \left( \displaystyle \sqrt{\kappa^L_\text{FES}}, 0 \right)$ &
 \parbox{0.9in}{$h_{11} = ( \varepsilon_L + \varepsilon_R + U_{LR}) \xi$} \\
 $y_{10,01} = 0$ &
 $\vec{e}_{10,01} = \left(
 -\displaystyle \sqrt{\kappa^{L}_\text{FES}},
 \sqrt{\kappa^R_\text{FES}} \right)$ \\
 $y_{00,11} = 0$ &
 $\vec{e}_{00,11} = \left(
 \displaystyle \sqrt{ \kappa^L_\text{FES}}, \sqrt{\kappa^R_\text{FES}} \right)$ \\
\end{tabular}
\end{ruledtabular}
\end{table}

We will now concentrate on the
Coulomb-blockade valley, i.e.,
$|\varepsilon_0|, \varepsilon_0+U \gg \Gamma_L, \Gamma_R$,
where $\varepsilon_0 = (\varepsilon_L + \varepsilon_R) / 2$,
and we assume a small level separation $\varepsilon_L - \varepsilon_R$
(see below).
In this regime only the singly-occupied states are important at low energies.
The RG flow is thus divided into three stages:
(i) In the first one, $\xi^{-1} \gg \max( |\varepsilon_0|, \varepsilon_0+U_{LR} )$,
and hence all the four filling configurations of the dots
must be treated on equal footing.
(ii) Then, as one enters the regime
$\min( |\varepsilon_0|, \varepsilon_0+U_{LR} ) \ll \xi^{-1} \ll
\max( |\varepsilon_0|, \varepsilon_0+U_{LR} )$,
the state with higher energy among the unoccupied and doubly-occupied
configurations becomes higher than the cutoff and is discarded;
(iii) Finally, for $\xi^{-1} \ll \min( |\varepsilon_0|, \varepsilon_0+U_{LR} )$,
only the singly-occupied states $10$ and $01$ are left.
In this last stage what remains is a Coulomb gas 
of only a single type of transitions.
It is thus equivalent to the one originally derived
by Anderson and Yuval for the single-channel anisotropic Kondo
model \cite{yuval_anderson},
indicating the equivalence of the two systems.
Under this mapping the two states $10$ and $01$ become, respectively,
the up and down states of the spin.
The main effect of the two first stages of the flow is to
establish the fugacity of the $10\rightleftharpoons01$ transition,
which is akin to the spin flip part ($J_{xy}$) of
the Kondo exchange coupling
(via virtual processes through the doubly occupied state
$11$ and the unoccupied state $00$).
In addition, these two first stages lead to renormalization of
the corresponding Coulomb-gas charge
(related to the $J_z$ part of the Kondo exchange)
and the energy difference between these states,
which is analogous to a local magnetic field $B_z$ along the $z$ axis
applied on the Kondo spin.

Comparing the two Coulomb gases we can extract the parameters of
the equivalent Kondo model.
To the leading order in
$\max(\Gamma^\ell_\text{FES})/\min(|\varepsilon_\ell|,\varepsilon_\ell+U_{LR})$
we find:
\begin{align}
 \rho_0 J_z = & 1 - \frac{\kappa^L_\text{FES} + \kappa^R_\text{FES}}{2}
    \nonumber \\ &
    + \sum_{\ell=L,R} { \frac{\kappa^\ell_\text{FES}
    \Gamma^\ell_\text{FES}}{\pi} \left\{
    \frac{Q_{2 \kappa^\ell_\text{FES}}
    (|\varepsilon_\ell| \xi)} {|\varepsilon_\ell|}
    + \frac{Q_{2 \kappa^\ell_\text{FES}} ( [ \varepsilon_\ell + U_{LR} ] \xi)}
    {\varepsilon_\ell + U_{LR}}
    \right\} }
    , \label{eqn:jz} \\
 \rho_0 J_{xy} = & \frac{2\sqrt{ \Gamma^L_\text{FES} \Gamma^R_\text{FES}} } {\pi}
    \nonumber \\ &
    \left[
    \frac{Q_{\kappa^L_\text{FES}+\kappa^R_\text{FES}}
        (|\varepsilon_0| \xi)} {|\varepsilon_0|}
    + \frac{Q_{\kappa^L_\text{FES}+\kappa^R_\text{FES}}
        ( [ \varepsilon_0 + U_{LR} ] \xi )}{\varepsilon_0 + U_{LR}}
    \right]
    ,  \label{eqn:jxy} \\
 B_z = & \varepsilon_L - \varepsilon_R 
    - \frac{\Gamma^L_\text{FES}}{\pi} \left[
    P_{2 \kappa^L_\text{FES}}(|\varepsilon_L| \xi)
    - P_{2 \kappa^L_\text{FES}} ( [ \varepsilon_L + U_{LR} ] \xi )
    \right]
    \nonumber \\ &
    + \frac{\Gamma^R_\text{FES}}{\pi} \left[
    P_{2 \kappa^R_\text{FES}} (|\varepsilon_R| \xi )
    - P_{2 \kappa^R_\text{FES}} ( [ \varepsilon_R + U_{LR} ] \xi )
    \right]
    ,  \label{eqn:hz}
\end{align}
where
$P_\mu(x) = \Gamma(1-\mu/2)/x^{1-\mu/2}$ with $\Gamma(z)$ the gamma function,
$Q_\mu(x) = (1-\mu/2)P_\mu(x)$,
and all the parameters of the original model refer to their bare values.
Without interactions (apart form the Coulomb coupling between the dots $U_{LR}$)
one has $\kappa^\ell_\text{FES} = 1$, and these expressions reduce to those obtained in
previous studies \cite{martinek03,lee07,kashcheyevs07,silvestrov07,kashcheyevs09,goldstein10}:
\begin{align}
 \label{eqn:jz0}
 \rho_0 J_z &= \frac{\Gamma_L}{\pi} \left(
    \frac{1}{\varepsilon_L + U_{LR}}
    + \frac{1} {|\varepsilon_L|}
    \right) 
    + \frac{\Gamma_R}{\pi} \left(
    \frac{1}{\varepsilon_R + U_{LR}}
    + \frac{1}{|\varepsilon_R|}
    \right) , \\
 \label{eqn:jxy0}
 \rho_0 J_{xy} &= \frac{2\sqrt{ \Gamma_L \Gamma_R} } {\pi}
    \left(
    \frac{1}{\varepsilon_0 + U_{LR}}
    + \frac{1}{|\varepsilon_0|}
    \right), \\
 \label{eqn:hz0}
 B_z &= \varepsilon_L - \varepsilon_R 
    - \frac{\Gamma_L}{\pi} \ln \frac{\varepsilon_L + U_{LR}}{|\varepsilon_L|}
    + \frac{\Gamma_R}{\pi} \ln \frac{\varepsilon_R + U_{LR}}{|\varepsilon_R|},
\end{align}
Since $U_{LR}$ is typically of the order of the bandwidth $\sim t_\ell$,
the $Q_\alpha(x)$ and $P_\alpha(x)$
functions only change the corresponding terms by factors of the order of unity
with respect to the case $\kappa^\ell_\text{FES} = 1$.
Hence, the main effect of having $\kappa^\ell_\text{FES} \ne 1$ thus
comes through the first term on the right hand side of Eq.~(\ref{eqn:jz}).

It should also be noted that
due to renormalization effects, $B_z$ depends not only on the
separation between the levels $\varepsilon_L - \varepsilon_R$,
but also on their average position $\varepsilon_0$.
In particular, for
$|\varepsilon_L-\varepsilon_R| < |\Gamma_L-\Gamma_R|$
the sign of $B_z$ will change as a common gate voltage
is varied across the Coulomb blockade valley.
Correspondingly, the sign of the average spin projection
$\langle S_z \rangle$
will also change, i.e., the two levels will swap their population
\cite{martinek03,lee07,kashcheyevs07,silvestrov07,kashcheyevs09,goldstein10}.

\begin{figure}
\includegraphics[width=7cm,height=!]{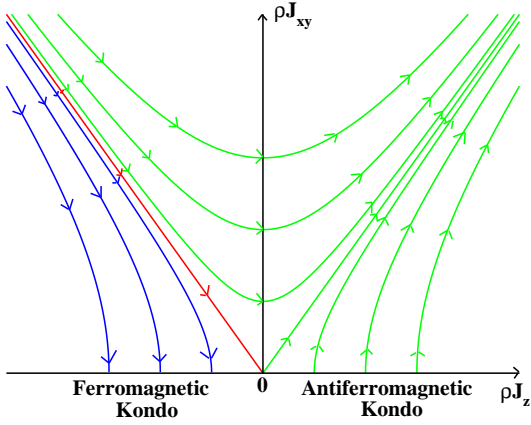}
\caption{\label{fig:kondo_rg}
(Color online)
Renormalization group flow of the anisotropic Kondo model
in the $J_{xy}$--$J_z$ plane.
In the isotropic ferromagnetic case
($J_z = -J_{xy}$, red straight line) and
all the region beneath it (blue lines) parameters flow
towards weak coupling, whereas
in the isotropic antiferromagnetic case
$J_z = J_{xy}$, green straight line) and
all the surrounding region (green lines) parameters flow
towards strong coupling.
}
\end{figure}

The famous Kondo renormalization group equations \cite{hewson,yuval_anderson},
\begin{align}
 \frac{d (\rho_0 J_{xy})}{d \ln \xi} & = (\rho_0 J_{xy}) (\rho_0 J_{z}), \\
 \frac{d (\rho_0 J_{z})}{d \ln \xi} & = (\rho_0 J_{xy})^2,
\end{align}
imply that (cf. Fig.~\ref{fig:kondo_rg})
for $J_z > - |J_{xy}|$
(including the case $U_\ell = 0$ and $U^\prime_\ell = 0$)
the exchange couplings grow under RG flow, and
we are in the strong-coupling (antiferromagnetic-like)
Kondo phase, where at low energies the Kondo spin is strongly-coupled
(entangled) into a singlet with the environment.
This behavior sets in at energies below the Kondo temperature,
which is parametrically smaller than any other energy scale in the problem,
and, in the generic anisotropic case
($J_z \gg J_{xy}$, where renormalization of $J_z$ is negligible),
scales with the exchange couplings as
$T_K \xi \sim (\rho_0 |J_{xy}|)^{1/(\rho_0 J_z)}$.
The average Kondo magnetization will be a smooth function of $B_z$
(i.e., the population is a continuous function of the gate voltage)
with the scale set by the Kondo temperature at low energies ($T \ll T_K$).
For $J_z < - |J_{xy}|$ (which may occur only if $\kappa^\ell_\text{FES}>1$)
we will enter the weak-coupling (ferromagnetic-like) Kondo regime,
in which $J_{xy}$ flows to zero under RG,
and the Kondo impurity becomes
effectively decoupled at low energies, so that the population will
be a discontinuous function of the gate voltage at zero temperature.
The transition between the strong-coupling and weak coupling regimes
(at $J_z = - |J_{xy}|$) is of the BKT type,
and the Kondo temperature goes to zero as $\ln(T_K) \sim -(J_z+|J_{xy}|)^{-1/2}$
when approaching the transition from the antiferromagnetic side.

This behavior can be understood in more physical terms.
For example, having 
$g_\ell \ne 1$ would shift $\rho J_z$ by $1-(1/g_L + 1/g_R)/2$.
By the Kondo renormalization group equations
this indicates that $J_{xy}$ processes involving tunneling out of one of the
leads and into the other are more (less) relevant
for $g_\ell>1$ ($g_\ell<1$), due to the enhanced (suppressed) tunneling
density of states to the endpoint
of a LL with attractive (repulsive) interactions \cite{giamarchi03,mattis74,kane92}.
In addition, having $U^\prime_\ell \ne 0$ modifies $J_z$.
The term linear in $U^\prime_\ell$ in the expression for $\kappa^\ell_\text{FES}$
represents the Mahan exciton effect \cite{noziers69,mahan},
and gives a positive (negative) contribution to $J_z$ when
$U^\prime_\ell > 0$ ($U^\prime_\ell < 0$):
when dot $\ell$ is occupied, then electrons are repelled from
(attracted to) the endpoint of lead $\ell$, thus enhancing (suppressing)
tunneling by the Pauli principle, and vice-versa.
The quadratic term in $U^\prime_\ell$ in the expression for $\kappa^\ell_\text{FES}$
represents the contribution of the Anderson orthogonality catastrophe \cite{noziers69,mahan},
which always suppresses transitions and hence causes a decrease in $J_z$.

In the following numerical calculations we will confine
ourselves to the case $U_L = U_R = U$ (so that $g_L = g_R = g$)
and $U^\prime_L = U^\prime_R = U^\prime$, i.e.,
$\kappa^L_\text{FES} = \kappa^R_\text{FES} = \kappa_\text{FES}$.
We will also take $N_L = N_R = N$.
For $\varepsilon_L=\varepsilon_R$ population switching will occur at
the point of particle-hole symmetry,
$\varepsilon_L=\varepsilon_R=-U_{LR}/2$.
The smooth-abrupt transition point will then be:
\begin{eqnarray}
\label{afes}
 \kappa^*_\text{FES} =
 \frac{1 + \frac{8}{\pi} \frac{\sqrt{ \Gamma^L_\text{FES}
\Gamma^R_\text{FES}} } {U_{LR}}
 Q_{2\kappa^*_\text{FES}}(U_{LR}\xi/2)}
 {1 - \frac{4}{\pi} \frac{\Gamma^L_\text{FES} + \Gamma^R_\text{FES}}{U_{LR}}
  Q_{2\kappa^*_\text{FES}}(U_{LR} \xi/2)},
\end{eqnarray}
These formulas are reliable for $\max(\Gamma^\ell_\text{FES}) \ll U_{LR}$, where
the critical $\kappa^*_\text{FES}$ is close to unity, so that
$Q_{2\kappa^*_\text{FES}}(U_{LR} \xi/2)$ is also almost equal to one,
and $\Gamma^\ell_\text{FES} \approx \Gamma_\ell$.
In addition, in the vicinity of the population switching we will get
\begin{equation}
B_z =
\frac{\Gamma(2-\kappa_\text{FES})}{(U_{LR} \xi/2)^{1-\kappa_\text{FES}}}
\frac{4(\Gamma^R_\text{FES}-\Gamma^L_\text{FES}) (\varepsilon_0+U_{LR}/2)} {\pi U_{LR}}.
\end{equation}
With the above expressions for $\kappa^*_\text{FES}$ one may find
the critical $U^*$ and/or $U^{\prime *}$ for any value of $\Gamma_\ell$ and $U_{LR}$.
One may also test if the critical value $U^*$ at $U^\prime=0$ and
the critical value of $U^{\prime *}$ at $U=0$ are compatible
(i.e., result in the same $\kappa^*_\text{FES}$)
even if $\Gamma_\ell$ are not small enough with respect to $U_{LR}$,
so that 
the critical $\kappa^*_\text{FES}$ itself
cannot be calculated analytically
(we still need to assume $\Gamma^\ell_\text{FES} \approx \Gamma_\ell$).


\section{Numerical Calculation of Entanglement Entropy}
\label{sec-num}

As discussed above, there are two possible phases for the
system. One corresponds to the Kondo anti-ferromagnetic phase, while
the other to the ferromagnetic phase of the Kondo model \cite{hewson}.

In order to calculate the EE
we shall cut the system depicted in Fig.~\ref{system}
into two
parts, L and R. The left region (L) includes the left lead and
the corresponding single-level QD (impurity),
while the right region (R) includes the right lead and
impurity. This is the most natural way to divide the system into two
equal parts and has the great advantage of being very natural in the
context of numerical DMRG calculations. Particle
transfer between the two regions is prohibited. Nevertheless, since the
two regions are coupled electrostatically,
they are correlated and possibly also entangled.
In the anti-ferromagnetic Kondo phase the ground state
of the system is a singlet, i.e.,
an equal superposition of two states (one in which
the left impurity is approximately full while the right one is approximately empty, and
vise-versa), resulting in two Schmidt coefficients $\alpha_1^2=1/2$
and $\alpha_2^2=1/2$, leading to
$S \equiv S_{L/R}=-\ln(1/2)=\ln(2)$.
On the other hand, for the ground state of the ferromagnetic Kondo phase,
the system is described by a single product many-body state,
resulting in $S=0$.
The transition
will thus
manifest itself in a
change of the EE between these two values.
In the following we will utilize this change
to facilitate the identification of
the critical transition point.

\subsection{Dot-Lead Interactions}

\begin{figure}
{\epsfxsize=3.5 in \epsffile{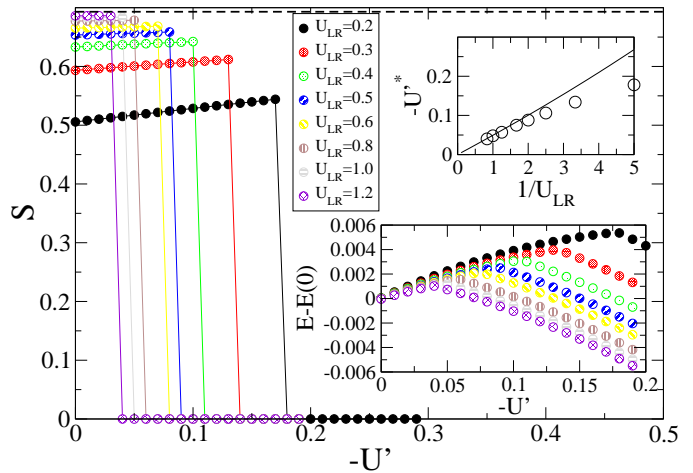}}
\caption{(Color online)
Numerical computation of the EE $S$ as function of the dot-lead
interactions $U^\prime$ for different values of $U_{LR}$,
the Coulomb interactions between the dots.
The leads are noninteracting ($U=0$).
The hopping matrix elements between each lead
and the corresponding dot are 
$t^\prime_L=0.1\sqrt{2}$ and $t^\prime_R=0.01\sqrt{2}$,
resulting in $\Gamma_L=0.02$ and $\Gamma_R=0.0002$.
In all cases each lead is of length $N=40$ (i.e., $40$ sites).
The transition between values of the EE close
to $\ln(2)$ (indicated by dashed line) and to zero is quite sharp,
and may be used to determine the critical value of $U^{\prime *}$.
Lower inset: The ground state energy $E(U^\prime)$ as function of
the dot-lead interaction. A cusp in the energy, indicative of a
level crossing, is clearly seen at values of $U^\prime$ corresponding
to the sharp drop in the EE.
Upper inset: The critical dot--lead interaction $U^{\prime *}$ as
function of the interdot interaction $U_{LR}$. The symbols correspond
to the numerical data while the line to the theoretical
predictions of Eq.~(\ref{udl}). Good agreement is observed
for large values of $U_{LR}$.
}
\label{fig_udl}
\end{figure}

We shall first consider the simplest case for which the leads are
a Fermi liquid (i.e., $U=0$ hence $g=1$) and the only interactions
are the interdot interactions $U_{LR}$ and the dot (impurity)-lead
interactions $U^\prime$.
From now on we take the value of the hopping matrix
element $t$ in the leads as our unit of energy.
We set $t^\prime_L=0.1\sqrt{2}$ and $t^\prime_R=0.01\sqrt{2}$,
resulting in $\Gamma_L=0.02$ and $\Gamma_R=0.0002$, and vary $U_{LR}$
and $U^\prime$.
Hereafter 
in this paper we tune the gate voltage 
to $\varepsilon_L=\varepsilon_R=-U_{LR}/2-0.0001$,
i.e., slightly above the expected
population switching point at half filling.
In all forthcoming calculations up to $320$ target states
were kept for the longer systems.
As detailed above, the mapping onto the Kondo model essentially depends
only on $\Gamma^L_\text{FES},\Gamma^R_\text{FES}$, and $\kappa_\text{FES}$, which, for the present
case, may be simplified to $\kappa_{FES}= [1-2 \tan^{-1}(U^\prime/2)/\pi]^2$.
As depicted in Fig.~\ref{fig_udl}, varying $U^\prime$ for a given value
of $U_{LR}$ will result in a modest increase in the EE as $U^\prime$ becomes
more negative, up to a point where a sharp drop in the EE occurs.
The larger the interdot interaction $U_{LR}$ is, the smaller the absolute
value of $U^\prime$ for which this drop occurs.
As can be seen in the inset, this drop in the EE occurs for the same
value of $U^\prime$ for which the ground state energy shows a downturn
typical of a crossing between two distinct ground-states, as expected from
a QPT.

According to the analysis in the previous section,
the transition
should occur at [see Eq.~(\ref{afes})]
\begin{eqnarray}
U^{\prime *}=2\tan\left\{\frac{\pi}{2}
\left[1-
\left(\frac{{\pi U_{LR}}+{{8\sqrt{\Gamma_L\Gamma_R}}
}}{{\pi U_{LR}}-{{4(\Gamma_L+\Gamma_R)}}}
\right)^{1/2}
\right]\right\}.
\label{udl}
\end{eqnarray}
Comparing this prediction depicted in the upper inset of Fig.~\ref{fig_udl} by the
curve to the numerical results 
for the transition corresponding to the
symbols, one can see that the results match for large values of $U_{LR}$, although
deviations appear at smaller values of $U_{LR}$. This is not surprising, since
in order to obtain the relation depicted in Eq.~(\ref{udl}) we assumed that
$\max(\Gamma_\ell)/U_{LR} \ll 1$.

\begin{figure}
{\epsfxsize=3.5 in \epsffile{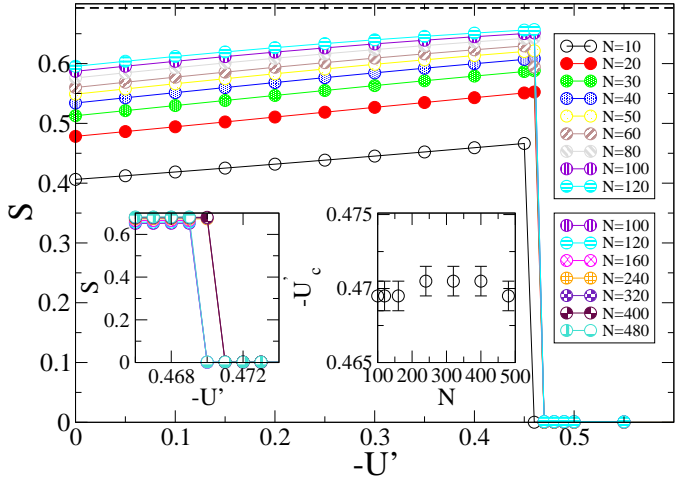}}
\caption{(Color online)
Numerical computation of the EE $S$ as function of the dot-lead
interactions $U^\prime$ for short
lead length $N \le 120$.
The leads are noninteracting ($U=0$).
Here $U_{LR}=1.2$, and
the hopping matrix elements between each lead
and the corresponding dot are 
$t^\prime_L=0.4\sqrt{2}$ and $t^\prime_R=0.03\sqrt{2}$,
resulting in $\Gamma_L=0.32$ and $\Gamma_R=0.0018$.
Again,
the transition between values of the EE close
to $\ln(2)$ (indicated by dashed line) and to zero is sharp,
and its location is
quite independent of the system size, resulting in
$U^{\prime}_c \approx -0.47$.
Left inset: A zoom in of the transition region
for larger system sizes $100 \le N \le 480$.
Right inset: the transition point $U^{\prime}_c$ as function of system size.
No substantial dependence of the critical dot-lead interaction on the
system size is apparent. Thus, the infinite length critical
dot-lead interaction is
$U^{\prime *} = -0.47 \pm 0.001$.
}
\label{fig_finite}
\end{figure}

The system size considered in Fig.~\ref{fig_udl} is rather small
($N_L = N_R = N=40$). Nevertheless, one would not expect strong finite-size effects
on the critical dot-lead interaction $U^{\prime *}$, since the mechanism by which tunneling is blocked (the Mahan exciton)
is local, i.e.,
attraction of an electron in the lead to the vicinity of the
occupied impurity due to the attractive lead-impurity coupling,
as explained in the previous section.
Indeed,
as can be seen in Fig.~\ref{fig_finite},
the main influence
of increasing $N$ is to bring the EE for 
$-U^{\prime}<-U^{\prime *}$
closer to
its expected infinite size value of $\ln(2)$,
while the estimation of the infinite length dot-lead critical
interaction $U^{\prime *}$
as well as the sharpness of the transition do not seem to dependend
on $N$. It is also interesting to note that $S$ increases
with $|U^\prime|$
for $-U^{\prime}<-U^{\prime *}$
(a trend which also appears in Fig.~\ref{fig_udl}).
This behavior 
could be understood as a result of the suppression of
the effective $\Gamma^\ell_\text{FES}$ by the dot-lead interaction
(cf.\ Table~\ref{tbl:fes}), consistent with the general trend
observed in Fig.~\ref{fig_udl}.

\subsection{Interactions in the Lead}

What happens if we consider a case with no dot-lead interaction
(i.e., $U^\prime=0$), but instead with nearest-neighbor interactions
in the leads ($U \ne 0$)? Since, as previously discussed, the
phase transition does not depend on the details
of the different interactions in the system, but rather on their
combined contribution to the Fermi edge singularity exponent
$\kappa_{\text{FES}}$, \cite{goldstein09,goldstein10b,goldstein10d} one expects that
knowing the critical value of
$U^{\prime *}$
one is able to predict the
critical value of $U^*$.
Thus, using Table~\ref{tbl:fes} and Eq.~(\ref{eqn:alpha_bethe}) one
can deduce that the critical LL parameter $g^*$ in the absence of
dot-lead interaction will be connected to the critical value
of $U^{\prime *}$ in the absence of interactions in the lead ($U=0$,
$g=1$) via $g^*=(1-2 \tan(U^{\prime *}/2)/\pi)^{-2}$,
from which we could find
$U^*$ employing Eq.~(\ref{gg}).
This relation should hold even
for $\Gamma_{L,R}/U_{LR} \sim 1$, for which Eq.~(\ref{udl}) no longer
holds.
Thus, for the parameters used 
in Fig.~\ref{fig_finite}, for which
$U^{\prime *} = -0.47$, one expects
$g^* \approx 0.75$,
resulting in
$U^{*} \approx 1$.

\begin{figure}
{\epsfxsize=3.5 in \epsffile{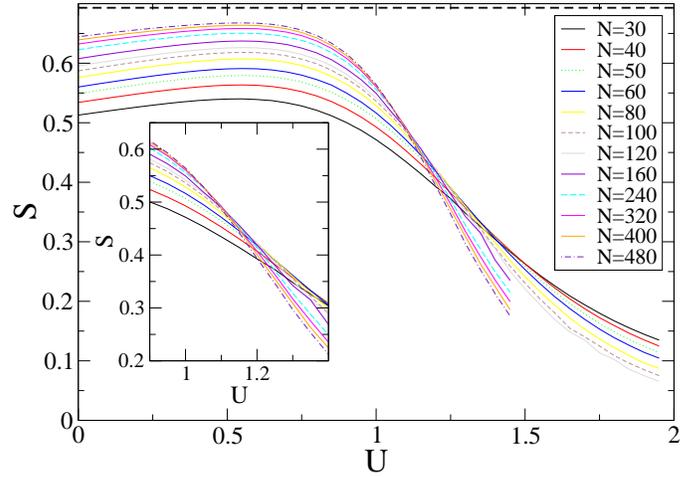}}
\caption{(Color online)
The numerically computed EE $S$ as function of the
interactions in the lead $U$ for different lengths $N$ of the leads.
The dot-lead interactions are absent ($U^\prime=0$).
The hopping matrix elements between each lead
and the corresponding dot are 
$t^\prime_L=0.4\sqrt{2}$ and $t^\prime_R=0.03\sqrt{2}$
($\Gamma_L=0.32$ and $\Gamma_R=0.0018$), and 
$U_{LR}=1.2$, corresponding to a Kondo
temperature of $T_K \approx 9 \cdot 10^{-5}$ for $U=0$.
\cite{goldstein10,martinek03,lee07,kashcheyevs07,silvestrov07,kashcheyevs09}
In contrast to the dot-lead interaction case
(Figs.~\ref{fig_udl}--\ref{fig_finite}),
the crossover between values of the EE close
to $\ln(2)$ (indicated by dashed line) and zero is rather broad.
As $N$ increases the values of $S$ for small $U$ increase, while they
decrease for larger values of the
interaction, resulting in crossings of the curves for
different values of $S$.
Inset: Unlike for second order phase transition in which all curves for
different $N$ cross at the same value, for the
BKT transition the crossing point shifts with $N$.
}
\label{fig_bkt}
\end{figure}

Indeed this can be demonstrated by the calculation of the EE
depicted in Fig.~\ref{fig_bkt}, where the finite-size behavior of
the same system considered in Fig.~\ref{fig_finite} (i.e., $U_{LR}=1.2$,
$t^\prime_L=0.4\sqrt{2}$, and $t^\prime_R=0.03\sqrt{2}$,
resulting in $\Gamma_L=0.32$ and $\Gamma_R=0.0018$)
is presented. A transition of $S$ from values close
to $\ln(2)$ to values approaching zero as $U$ increases is
apparent. Nevertheless, the transition is much more gradual than
for the dot-lead interaction (see Fig.~\ref{fig_finite}),
which makes sense since the LL character of the lead will be fully
developed only for large sample sizes.
A crossing between the curves
corresponding to different sizes is apparent. While for small values of
$U$ larger system sizes correspond to larger values of $S$,
the opposite occurs for large values of $U$, which is a hallmark
of finite-size scaling of phase transitions \cite{barber83}.

It is important to emphasize that since we are dealing with
a BKT phase transition, the
finite-size scaling differs from that of a
traditional second order phase transition \cite{barber83}.
While for a second order transition all the curves are
expected to cross at the same point, for a BKT transition
the crossing point will drift as function of size and
only at the limit of $N \rightarrow \infty$ will the crossing
point correspond to the critical value $U^*$ of the lead
nearest-neighbor interaction. The drift in the crossing point
can be seen clearly in the inset to Fig.~\ref{fig_bkt}, where
the crossing between the $N=30$ and $N=40$ curves occurs at
$U_c \approx 1.5$, while between $N=400$ and $N=480$ $U_c \approx 1.05$.

Unlike for finite size scaling for second order transitions, there is no
consensus on the optimal method to extract the transition
point $U^*$. Therefore, we shall employ three different
methods to identify the critical interaction for which the
transition occurs: (I) extrapolation of the crossing point at
$N \rightarrow \infty$ in the spirit of the phenomenological renormalization
group (PRG) procedure \cite{barber83,chen03}; (II) extrapolation
of the transition point using a scaling ansatz inspired
by the homogeneity condition method \cite{roncaglia08};
(III) identification of the transition point by an heuristic
scaling function. All these procedures give a similar 
estimate
of $U^*$, which is in good agreement with the correspondence
between $U^*$ and $U^{\prime *}$.

\begin{figure}
{\epsfxsize=3.5 in \epsffile{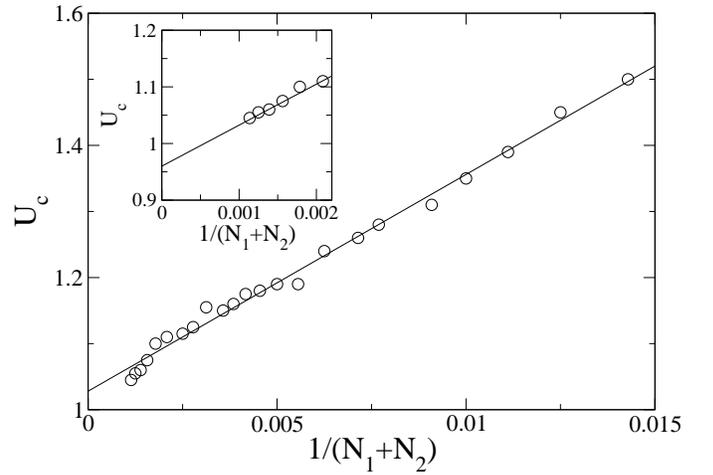}}
\caption{Obtaining the critical nearest-neighbor interaction
in the lead $U^*$ by way of the extrapolation Eq.~(\ref{uc}).
The symbols correspond to the crossing point between
the curves in Fig.~\ref{fig_bkt} for given values of $N_1$ and $N_2$.
Here we consider ($N_1$,$N_2$) pairs
that correspond to successive and
next-to-successive values of $N$ appearing in Fig.~\ref{fig_bkt}.
The line represents 
a fit to Eq.~(\ref{uc}).
Inset: A fit to Eq.~(\ref{uc}) performed only for the large $N$ values.
}
\label{fig_prg}
\end{figure}

For the PRG inspired (for details
of the PRG procedure see Ref. \onlinecite{barber83})
extrapolation of the crossing point as 
$N \rightarrow \infty$, one defines $U_c(N_1,N_2)$, which is
the value of the crossing between the curve corresponding to
length $N_1$ and the curve corresponding to $N_2$, where $N_1$ and $N_2$
are two successive (or next to successive) length values.
We then extrapolate the behavior of the crossing points by
the formula \cite{chen03}
\begin{eqnarray}
U_c (N_1,N_2) = U^*+\frac{C}{N_1+N_2},
\label{uc}
\end{eqnarray}
where $C$ is a constant, and the value of the crossing point at infinite
length is assumed to correspond to the transition point $U^*$.
As can be seen in Fig.~\ref{fig_prg}, Eq.~(\ref{uc}) works well for the
entire range leading to $U^* = 1.02$. 
When one examines more carefully
the large $N$ region (see the inset) the estimation of the critical
point shifts a bit to $U^* = 0.96$. Both values are consistent with
the expectations based on the dot-lead interaction data
($U^*=1 \pm 0.1$).

\begin{figure}
{\epsfxsize=3.5 in \epsffile{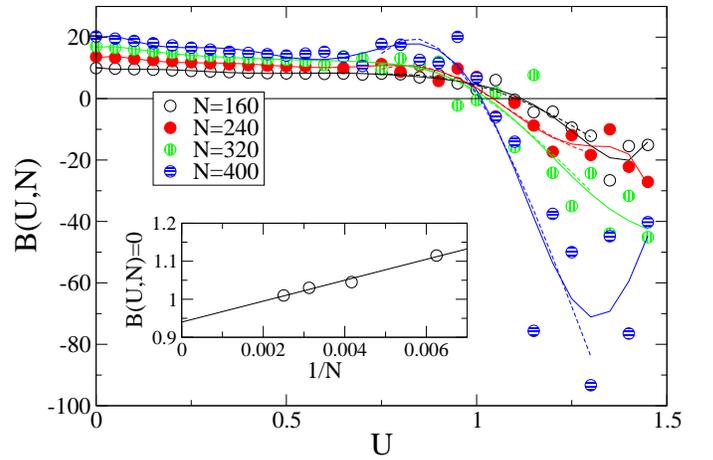}}
\caption{(Color online)
$B(U,N)$ [defined in Eq.~(\ref{hcm})] as function
of the lead nearest-neighbor interaction $U$, for different
system lengths $N$. The symbols correspond to the numerical
results while the continuous curves to a fit using a
6th degree polynomial and the dashed curve to a fit using
a 3rd degree polynomial in the crossing region}.
Inset: extrapolating the critical value $U^*$ for $N \rightarrow \infty$.
The symbols depict the values for which $B(U,N)=0$, deduced from
the curves appearing in the main figure.
\label{fig_b}
\end{figure}

The second method is inspired by the homogeneity condition method
proposed and described in detail
in Ref.~\onlinecite{roncaglia08}.
In this method a function $b(U,N)$ is constructed from the expectation
value of the term in the Hamiltonian driving the transition (in our case
the interactions in the lead). Then the transition point is determined
by the condition $B(U^*,N)=\partial_N[N^3 \partial_N b(U^*,N)]=0$.
Here we shall
replace the expectation value $b(U,N)$ by the EE $S(U,N)$,
and rewrite the function $B(U,N)$
as a discrete differentiation:
\begin{eqnarray}
B(U,N)=N^3 S^{\prime \prime}(U,N)+[3 N^2 - (\delta N)^2]S^{\prime}(U,N),
\label{hcm}
\end{eqnarray}
where $S^{\prime}$ ($S^{\prime \prime}$)
is the first (second) order discrete differentiation
of $S(U,N)$ \{i.e.,
$S^{\prime}(U,N)=[S(U,N+\delta N)-S(U_l,N-\delta N)]/(2 \delta N)$ and
$S^{\prime \prime}(U,N)=[S(U,N+\delta N) + S(U,N-\delta N) - 2S(U,N)]/(\delta N)^2$\},
and $\delta N$ is the differentiation step.
In Fig.~\ref{fig_b} $B(U,N)$ is plotted for the larger system
sizes ($N=160$, $240$, $320$, $400$) with $\delta N = 80$. Since for
higher values of $U$ there is strong scatter in the data
(as indeed has been noticed in other applications of the HMC, see
cf. Ref. \onlinecite{roncaglia08}), we have
interpolated $B(U,N)$ over the whole range
by a fit to a 6th degree polynomial (continuous curves),
and in the vicinity
of $B(U,N)=0$ by a fit to a cubic polynomial (dashed curve).
The transition point for each size is determined by
$B(U,N)=0$, i.e., the point in which the curve crosses the x-axis.
This point is found to be insensitive to the degree of the interpolation polynomial used.
Extrapolating the critical value $U^*$ from the finite size
values corresponding to $B(U,N)=0$ (see inset of Fig.~\ref{fig_b})
results in $U^* = 0.94$.

\begin{figure}
{\epsfxsize=3.5 in \epsffile{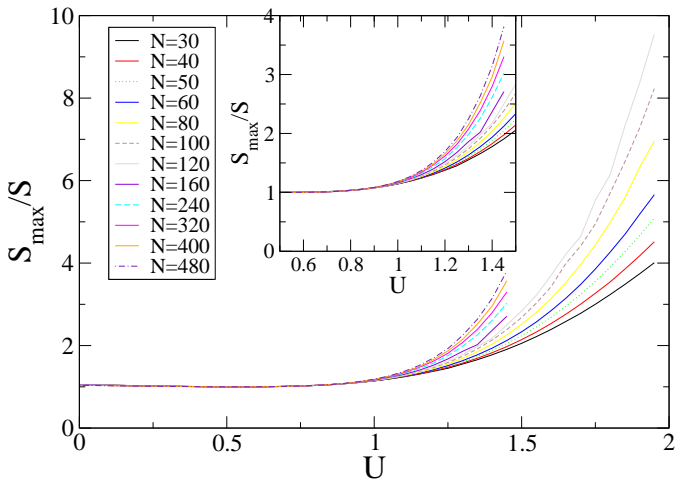}}
\caption{(Color online)
Heuristic scaling [cf.\ Eq.~(\ref{sscaling})]
of the different curves appearing
in Fig.~\ref{fig_bkt}. The curves depict the scaling
function $\tilde S(U) = S_\text{max}(N)/S(U,N)$ for different system
sizes $N$ as function of $U$.
Inset: Zoom into the transition region. Up to $U \approx 0.95$
all the curves coalesce. For larger values the curves begin to separate.
We identify this separation point as the critical $U^*$.}
\label{fig_s}
\end{figure}

The third method we shall use in order to determine the critical
value of $U^*$ involves a 
heuristic scaling function of the
different curves appearing in Fig.~\ref{fig_bkt}.
This scaling is based on the observation that the maximum entropy for
all system sizes seems to appear at the same value of $U \approx 0.55$,
and that the general form of $S(U,N)$ around this value shifts
in a rigid manner as function of $N$. Thus, we postulate that 
for $U<U^*$ one may collapse all the different
curves onto a single curve by the following heuristic scaling function:
\begin{eqnarray}
\tilde S(U)=\frac{S_\text{max}(N)}{S(U,N)},
\label{sscaling}
\end{eqnarray}
where $S_\text{max}(N)=S(U=0.55,N)$ is the maximum value of the EE for
a given length $N$ of the system. As can be seen in Fig.~\ref{fig_s},
applying this heuristic function collapses all the different curves on a single
one up-to a certain value of $U$, while for larger values the
curves diverge. This is somewhat
similar to the situation one encounters for the
PRG scaling of the energy gap in the BKT transition \cite{barber83}.
Identifying the point for which the curves begin to diverge
with the critical point (as is done in the PRG procedure) results
in $U^* \approx 0.95$, in agreement with the other methods we employed here
for identifying the critical point.

\subsection{Both Dot-Lead and Intra-Lead Interactions}

Now we shall discuss the behavior for the case where both $U^\prime$ and $U$
are nonzero. 
We will consider the case where both $U^\prime>0$ and $U>0$,
which is the case expected to be encountered in realistic experimental
devices. We shall set the dot-lead tunneling matrix elements
$t^\prime_L=0.2\sqrt{2}$ and $t^\prime_R=0.02\sqrt{2}$, corresponding to
$\Gamma_L=0.08$ and $\Gamma_R=0.0008$, while retaining $U_{LR}=1.2$.
For these parameters $\max(\Gamma_\ell)/U_{LR} \ll 1$ and we expect the
determination of the
critical point which depends on the parameter $\kappa^*_{\text{FES}}$
given in Eq.~(\ref{afes}) to hold. Setting $U^\prime=0.3$ and using
the relationship given in Eq.~(\ref{eqn:alpha_bethe}), the transition
is expected to occur at $U^*$ obeying the following relation:
\begin{multline}
U^\prime=\sqrt{4-(U^*)^2} \times \\
\tan \left[ \cos^{-1}\left(-\frac{U^*}{2}\right)
\left( 1-\sqrt{\frac{\pi \kappa^*_\text{FES}}{2 \cos(-U^*/2)}} \right) \right].
\label{rel}
\end{multline}
The right hand side, for a given value of $\Gamma_L$, $\Gamma_R$, and
$U_{LR}$, depends only on $U^*$. Thus by plotting the right side of Eq.~(\ref{rel}) and determining where it crosses the value of $U^\prime$,
it is possible to evaluate $U^*$. For the parameters considered here,
the critical value corresponds to $U^*=0.96$. There is an additional
crossing at $U^*=1.95$, but since this is very close to the LL-charge
density wave phase transition point (at $U=2$) it would be extremely
hard to observe 
it numerically for reasonable system sizes.

One may wonder why in the previous sub-section we did not consider
also the case of $\max(\Gamma_\ell)/U_{LR} \ll 1$, which
would facilitate a direct comparison between $U^*$ and theory, instead
of discussing a case for which $\Gamma_\ell$ are comparable to $U_{LR}$, leaving
only the possibility to compare between the numerically computed
$U^{\prime *}$ and $U^*$? The reason is that for $\max(\Gamma_\ell)/U_{LR} \ll 1$ and $U^\prime=0$,
$U^*$ tends to be close to zero, i.e., $g \approx 1$. For
values of the LL close to the non-interacting case, one must go to very large
systems to see the LL behavior developing \cite{weiss07}, which is beyond our
current capabilities.

\begin{figure}
{\epsfxsize=3.2 in \epsffile{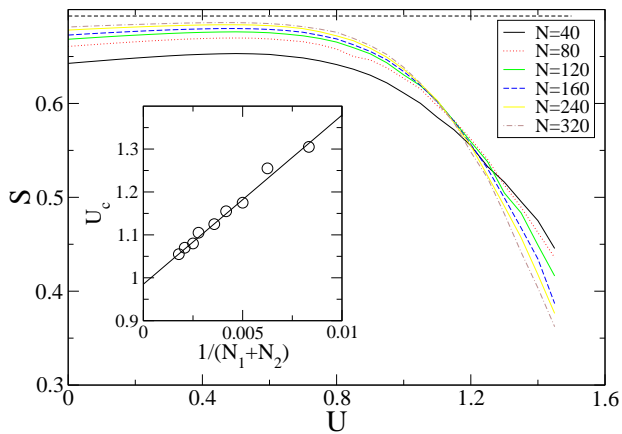}}
\caption{(Color online)
The numerically computed EE $S$ as function of the
interactions in the lead $U$, for a constant dot-lead
interaction $U^\prime=0.3$ and different lengths $N$ of the lead.
The hopping matrix elements between each lead
and the corresponding dot are given by
$t^\prime_L=0.2\sqrt{2}$ and $t^\prime_R=0.02\sqrt{2}$
($\Gamma_L=0.08$ and $\Gamma_R=0.0008$), 
while $U_{LR}=1.2$, corresponding to a Kondo
temperature of
$T_K \approx 10^{-13}$ at $U=0$.
\cite{goldstein10,martinek03,lee07,kashcheyevs07,silvestrov07,kashcheyevs09}
The
transition between values of the EE close
to $\ln(2)$ (indicated by dashed line) and to zero is steeper than
in Fig.~\ref{fig_bkt}, although the typical BKT transition crossing
are still clearly evident.
Inset: An extrapolation of the critical $U^*$ in the
same manner as in Fig.~\ref{fig_prg}.
The symbols depict the crossing point between
the curves in the main figure corresponding to $N_1$ and $N_2$,
where successive and next to successive lengths are taken.
The line represents a fit to Eq.~(\ref{uc}).
}
\label{fig_combined}
\end{figure}

Returning to the case at hand, it can be clearly seen
in Fig.~\ref{fig_combined} that as in the
previous cases, a typical BKT crossing of the $S$ as function of $U$
for the different lengths is observed. 
Using the
extrapolation of the crossing point at
$N \rightarrow \infty$ following the PRG procedure, and fitting
it to Eq.~(\ref{uc}) (see inset), an extrapolated value of $U^*=0.98$
is obtained, in good agreement with the analytical prediction.

\section{Discussion}
\label{sec-dis}

We have shown that in the presence of repulsive
interaction in the lead
($U>0$, $g<1$), the system depicted in Fig.~\ref{system} may show an abrupt
population switching, i.e., an abrupt swap of the
left dot-right dot population
as function of the applied gate voltage.
The nature of the population switching has a clear signature
in the behavior of the EE. For smooth switching the system
can be mapped on the antiferromagnetic Kondo model, resulting in
a finite entanglement between the left and right sub-systems.
On the other hand, for abrupt switching the system corresponds to
a ferromagnetic Kondo model, for which there is no entanglement
between the sub-systems.

Using this behavior of the EE, and the fact that it lends itself to
straightforward calculation within
the framework of numerical DMRG, we were able to use
finite size scaling to identify the QPT between the smooth and abrupt
switching phases. 
We have found that
using the signature of the QPT on the EE behavior indeed
gives an accurate method to study the transition properties. This
reproduces the analytical results obtained by mapping the the system
onto a Kondo model. One may expect that the EE could be used in a similar
fashion to identify and study different QPTs related to dynamical
impurities. \cite{vojta06,lehur08,kane92,furusaki93,lerner08,goldstein10a,goldstein10c,elste10,
furusaki02,sade05,lehur05,weiss0607,wachter07,weiss08,fiete0810,goldstein09,goldstein10b,hamamoto10,mora10,goldstein10d,impurity_qpt}

Finally, it should be noted 
that according to
Eq.~(\ref{afes}), in the absence of dot-lead interactions ($U^\prime=0$),
and for strong interdot interactions $U_{LR} \gg \max(\Gamma_\ell)$, 
the abrupt population switching will occur at
values of the LL parameter
smaller than a critical value which is quite close to unity
\begin{eqnarray}
1 - g^* \sim \frac {\max(\Gamma_\ell)}{U_{LR}}. 
\label{glimit}
\end{eqnarray}
Thus, for large interdot interaction
$U_{LR} \gg \max{\Gamma_\ell}$
, even a weak interaction
in the leads will result in an abrupt population switching. Once
dot-lead interactions are taken in account,
the critical $g^*$ will be further decreased by $2 U^\prime/(\pi t)$.
Since $t$ is essentially the
band width and the dot-lead interaction is parametrically much smaller,
the abrupt population switching
phase will appear for values of $g \approx 1$
even when repulsive dot-lead-interactions are included,
and should thus be accessible experimentally.
The population switching can be probed using a
quantum point contact as a charge sensor.\cite{charge_sensing}

\begin{acknowledgments}
We would like to thank D. I. Golosov, A. Schiller, and J. von Delft
for useful discussions.
M. G. is supported by the Adams Foundation
of the Israel Academy of Sciences and Humanities,
the Simons Foundation, the Fulbright Foundation, and the
BIKURA (FIRST) program of the Israel Science Foundation.
Financial support from the Israel Science Foundation (Grants 569/07
and 686/10), SPP 1285 ``Spintronics'', and Israel-Russia MOST grant is
gratefully acknowledged.
\end{acknowledgments}


\begin{references}
\bibitem{uu} For $U_\ell/t_\ell>2$ a charge density wave is formed,
while for $U_\ell/t_\ell<-2$
phase separation into dense and sparse regions will take place.
\end{references}
\end{document}